\documentclass[twocolumn, showpacs, prl,unsortedaddress]{revtex4}

\usepackage{graphicx}

\usepackage{dcolumn}

\usepackage{color}
\usepackage{bm}

\begin{document}

\title{Contrast between spin and valley degrees of freedom}

\date{\today}

\author{T.\ Gokmen}

\author{Medini\ Padmanabhan}

\author{M.\ Shayegan}

\affiliation{Department of Electrical Engineering, Princeton
University, Princeton, NJ 08544}

\begin{abstract}

We measure the renormalized effective mass ($m^*$) of interacting
two-dimensional electrons confined to an AlAs quantum well while we control their distribution between two spin and two valley subbands. We observe a
marked contrast between the spin and valley degrees of freedom: When electrons
occupy two spin subbands, $m^*$ strongly depends on the valley occupation,
but not vice versa. Combining our $m^*$ data with the measured
spin and valley susceptibilities, we find that the renormalized
effective Lande g-factor strongly depends on valley occupation, but the renormalized conduction-band deformation potential is nearly independent of the spin occupation.
\end{abstract}


\maketitle

Low-disorder two-dimensional electrons provide a nearly ideal system for the
study of electron-electron interaction. The interaction
strength, characterized by $r_s=1/\sqrt{\pi n}a_B^*$, the average
inter-electron spacing measured in units of the effective Bohr
radius, can easily be tuned by varying the density of an interacting
two-dimensional electron system (2DES). In the Fermi liquid theory, electron-electron interaction renormalizes the fundamental
parameters of the 2DES, such as the effective mass ($m^*$) and the spin susceptibility ($\chi^*_s$
${\propto}$ $g^{*} m^{*}$), where $g^*$ is the effective
Lande g-factor \cite{CeperleyPRB1989}. In particular, $\chi^*_s$ and $m^*$ are expected to
be larger than the band values ($\chi_{s,b}$ and $m_b$) for large
$r_{s}$ \cite{CeperleyPRB1989, AttaccalitePRL2002, AsgariPRB2006, AsgariSSC2004,
DePaloPRL2005, GangadharaiahPRL05, ZhangPRB05}. Enhancements of $\chi^*_s$ and $m^*$ at large
$r_{s}$ are indeed observed in a number of different 2DESs \cite{PanPRB99,
PudalovPRL02, ShashkinPRB02, ShashkinPRL03, VakiliPRL04,
ShkolnikovPRL04, TanPRL2005, FangPR1968, OkamotoPRL1999,
VitkalovPRL2001, ShashkinPRL2001, ZhuPRL2003, PrusPRB2003,
TutucPRB2003, GunawanPRL06}.

In addition to $r_s$, the role of spin and valley degrees of freedom on $\chi^*_s$ and $m^*$
renormalization has been explored both experimentally \cite{ShashkinPRL03, ShkolnikovPRL04, GunawanPRL06, PadmanabhanPRL08, GokmenPRL08, GokmenPRB09} and theoretically \cite{GangadharaiahPRL05, ZhangPRB05, MarchiPRB09, DasSarmaPRB09}. Measurements of $\chi^*_s$ in AlAs 2DESs \cite{ShkolnikovPRL04,GunawanPRL06}, e.g., revealed that $\chi^*_s$ is
smaller for a two-valley system than it is for a single-valley
system. This unexpected result was subsequently explained by theoretical studies \cite{ZhangPRB05, MarchiPRB09, DasSarmaPRB09}. The valley susceptibility ($\chi^*_v$ ${\propto}$ $E^{*}_2 m^{*}$) of AlAs 2DESs, defined as the rate of valley polarization with applied strain (in analogy to $\chi^*_s$ which is defined as the rate of spin polarization with applied magnetic field), has also been measured \cite{GunawanPRL06} ($E^*_2$ is the conduction band
deformation potential). It was found that $\chi^*_v$ depends on the spin
subband occupation \cite{PadmanabhanPRB08} in a similar way that
$\chi^*_s$ depends on the valley occupation. This observation is consistent with the expectation from theories which treat spin and valley as equivalent degrees of freedom \cite{MarchiPRB09,DasSarmaPRB09}.

In this Letter, we report measurements of $m^*$, $\chi^*_s$, and $\chi^*_v$ for an interacting
2DES confined to a wide AlAs quantum well as a function of $r_s$. The data reveal that the spin and valley degrees of freedom are not equivalent in this system. In our samples we can tune the spin and valley energies so that all four possible spin
and valley subband occupations are realized:
$s_2v_2$, $s_2v_1$, $s_1v_2$ and $s_1v_1$, where $s$ and $v$ stand for
spin and valley, and 1 and 2 denote the number of occupied spin or valley
subbands \cite{PolarizationNote}. Our results, summarized in Figs. 1 and 2, illustrate an intriguing contrast between the role of spin and valley degrees of freedom in $m^*$, $g^*$ and $E_2^*$ renormalization:

\begin{enumerate}
\item As seen in Fig. 1, for a system where electrons reside in two spin subbands, $m^*$
for the two-valley case is larger than $m^*$ for the
single-valley case. However, when the electrons reside in two valleys, $m^*$ depends only slightly on the
spin occupation. In other words, $m^*_{s_1v_2} > m^*_{s_2v_1}$.

\item As reported before \cite{ShkolnikovPRL04, GunawanPRL06, PadmanabhanPRB08}, the dependence of $\chi^*_s$
${\propto}$ $g^{*} m^{*}$ on valley occupation is similar to the dependence
of $\chi^*_v$ ${\propto}$ $E^{*}_2 m^{*}$ on spin subband occupation, namely, $\chi^*_s$ for $v_1$ is larger than for $v_2$, and $\chi^*_v$ for $s_1$ is larger than for $s_2$ (Fig. 2). Combining our $m^*$ data with the measurements of $g^*m^*$ and $E_2^*m^*$ done in the same system, we also deduce values for $g^*$ and $E_2^*$ for different valley and spin occupations (Fig. 2). Deduced $g^*$ values are smaller for the $v_2$ case compared to the $v_1$ case. However, the deduced $E_2^*$ values
are independent of the spin occupation, i.e., they are the same for $s_2$ and $s_1$.
\end{enumerate}

We studied high-mobility 2DESs confined to modulation-doped AlAs quantum wells grown by molecular beam epitaxy on a
(001) GaAs substrate \cite{ShayeganPSS06}. The AlAs well width in our specimens ranged from 11 to 15 nm. In these samples, the
2D electrons occupy two energetically degenerate conduction-band
valleys with elliptical Fermi contours,
each centered at an X point of the Brillouin zone and with an
anisotropic mass (longitudinal mass $m_{l}=1.05$ and transverse
mass $m_{t}=0.20$, in units of free electron mass, $m_{e}$) \cite{ShayeganPSS06}. The band mass in our 2DES has a value $m_b=\sqrt{m_lm_t}=0.46$, the band g-factor is $g_b=2$, the band value for the deformation potential is $E_{2,b}=5.8$ eV. The degeneracy between the valleys can be lifted by applying a symmetry breaking strain in the plane \cite{ShayeganPSS06}, allowing us to tune the valley occupation \textit{in situ}. Moreover, we control the spin occupation via the application of magnetic field. The
magneto-resistance measurements were performed in a $^3$He system with a base temperature of 0.3 K and equipped
with a tilting stage, allowing us to vary the angle between the
sample normal and the magnetic field \textit{in
situ} in order to tune the Zeeman energy at a fixed perpendicular magnetic filed.

Figure 1 summarizes the results of our $m^{*}$ measurements as a function of density for different valley and spin occupations. We deduce $m^{*}$ via analyzing the temperature dependence of the
strength of the Shubnikov-de Haas oscillations using the standard Dingle
expression. Details of our analysis are published in Refs.
\cite{PadmanabhanPRL08, GokmenPRL08, GokmenPRB09}. For each density, we carefully tune the valley and spin splitting energies via applying appropriate amounts of strain and in-plane magnetic field so that the Fermi energy at a given perpendicular magnetic field is in the gap between two energy levels separated by the cyclotron energy, and then measure the amplitude of the resistance oscillation as a function of temperature.

For the $s_1v_1$ case the measured $m^*$ is smaller than the band value in the entire density range of our experiments \cite{PadmanabhanPRL08}. This unexpected suppression of $m^*$ for a fully spin and valley polarized 2DES is also observed in narrow AlAs 2DESs \cite{GokmenPRB09}, and was very recently reproduced theoretically \cite{AsgariPRB09, DrummondPRB09}. For all other combinations of spin and valley occupations, $m^*$ increases with increasing $r_s$. This is the trend observed in other 2DESs where either two valley or spins are occupied \cite{PanPRB99, PudalovPRL02, ShashkinPRB02, ShashkinPRL03, VakiliPRL04, GokmenPRB09, TanPRL2005}. The highlight of our work is the contrast seen in Fig. 1 between the data for $s_1v_2$ and $s_2v_1$ cases: \textit{$m^*$ is much larger for $s_1v_2$ compared to $s_2v_1$}. Moreover, the data reveal that when two valleys are occupied, $m^*$ shows only a slight dependence on the spin occupation, whereas for $s_2$, $m^*$ for $v_1$ is much smaller than it is for $v_2$.

The role of spin polarization on $m^*$ renormalization has been addressed theoretically \cite{GangadharaiahPRL05}, and it was concluded that $m^*$ is independent of the spin polarization for a valley degenerate ($v_2$) 2DES. This conclusion
is in agreement with our data ($m^*$ depends only slightly on $s$ for $v_2$) and also with the data for 2DESs in Si-MOSFETs \cite{ShashkinPRL03}. If the valley and spin degrees of freedom were identical, one would expect similar $m^*$ values when $s$ and $v$ are interchanged, meaning that $m^*$ should not
depend on valley occupation when two spins are occupied. Our data of Fig. 1 clearly contradict this expectation: $m^*_{s_2v_1}$ is 30\% to 40\% smaller than $m^*_{s_2v_2}$. Evidently in our 2DES the spin and valley indices are not alike in the
determination of $m^*$.

\begin{figure}
\centering
\includegraphics[scale=0.57]{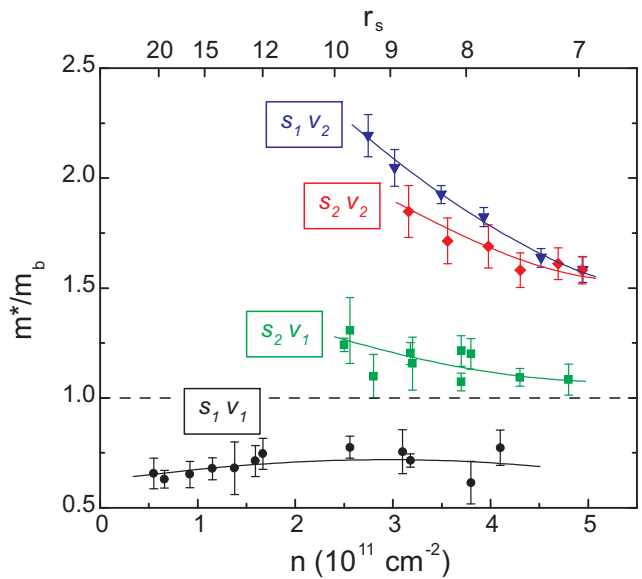}
\caption{(Color online) Density dependence of effective mass, $m^*$,
normalized to the band value, for 2D elecrons confined to wide AlAs quantum wells. Data are shown for four possible spin and valley occupations: $s_2v_2$, $s_2v_1$, $s_1v_2$ and $s_1v_1$, where $s$ and $v$ stand for spin and valley, and 1 and 2 denote
the number of spin/valley subbands that are occupied. The curves through the data points are guides to the
eye.}
\end{figure}

\begin{figure*} \centering
\includegraphics[scale=0.29]{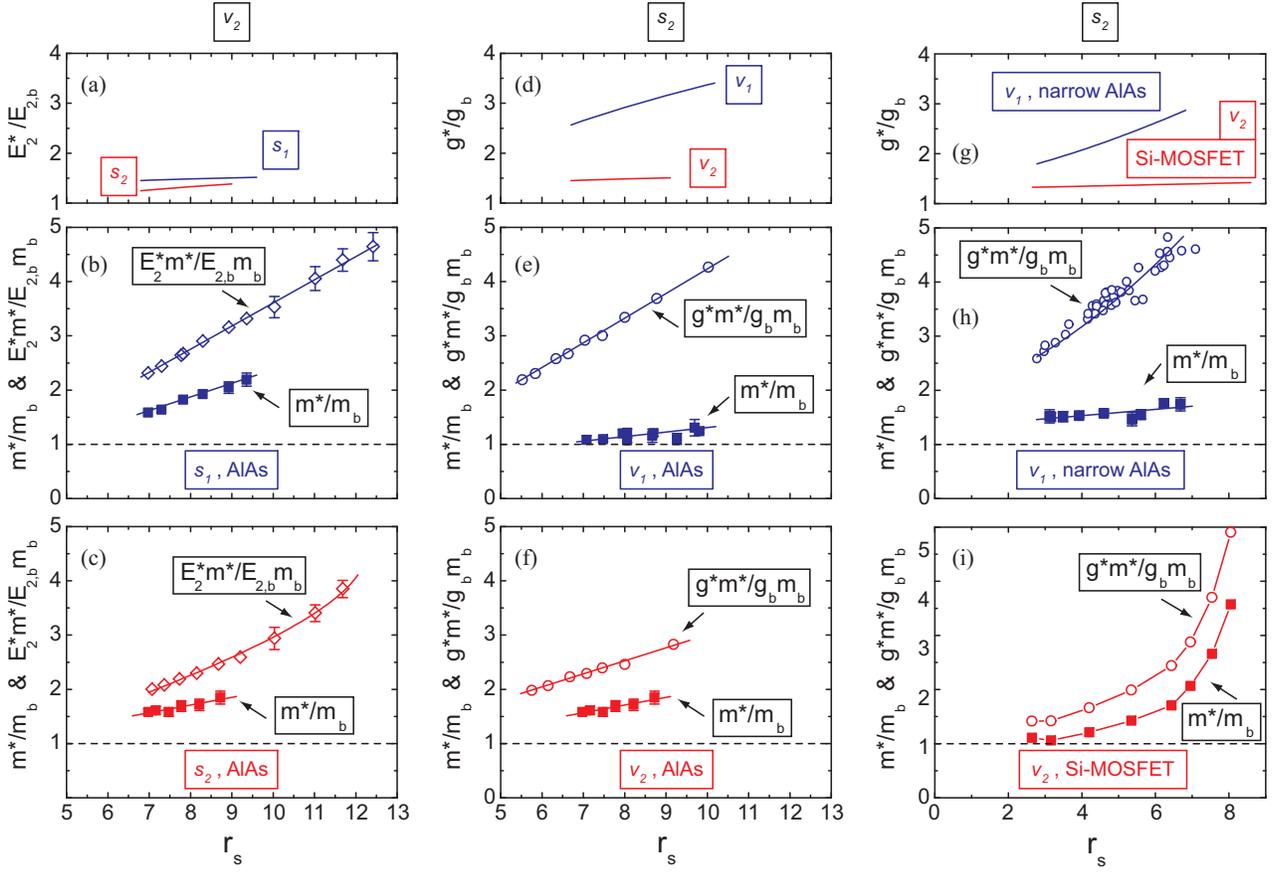}
\caption{(Color online) Effective mass ($m^*$, squares), valley susceptibility ($E_2^*m^*$, diamonds), and spin susceptibility ($g^*m^*$, circles) are shown as a function of $r_s$ in the lower six panels. The upper three panels show $E_2^*$ and $g^*$ deduced from the $m^*$ and susceptibility data. The left and central panels contain data for wide AlAs quantum wells for the different combinations of valley and spin occupations as indicated, e.g., in (b) data for $s_1v_2$ are shown. Note that in wide AlAs quantum wells the electrons occupy one or two in-plane valleys with \textit{anisotropic} Fermi contours. The right panels present data for 2D electrons in Si-MOSFETs (Ref. \cite{ShashkinPRB02}) where two out-of-plane valleys are occupied, and in narrow AlAs quantum wells (Refs. \cite{VakiliPRL04, GokmenPRB09} where the electrons occupy a single out-of-plane valley; note that these out-of-plane valleys have an \textit{isotropic} Fermi contour.}
\end{figure*}


Next we discuss the contrast between spin and valley degrees of
freedom that is revealed in the measurements of $g^*$ and $E_2^*$.
The values of spin and valley susceptibilities, $\chi^*_s$
${\propto}$ $g^*m^*$ and $\chi^*_v$ ${\propto}$ $E_2^*m^*$, were measured as a function of $r_s$ and at different valley and spin subband occupations via "coincidence" measurements \cite{ShkolnikovPRL04, GunawanPRL06, PadmanabhanPRB08}. In such measurements, the valley and spin splitting energies are tuned very carefully so that two Landau levels corresponding to different spins or valleys coincide at the Fermi energy. The coincidence is signaled by a maximum in the resistance at integer filling factors where, in the absence of the coincidence, a minimum is expected as the Fermi energy resides in a gap between two energy levels. From the values of strain and tilt angle at which such coincidences occur, we deduce the Zeeman- and/or the valley-splitting energies normalized to the cyclotron energy. These energies directly give $g^*m^*$ and $E_2^*m^*$. Combining the measured $g^*m^*$ and $E_2^*m^*$ with the $m^*$ data, we deduce values for $g^*$ and $E_2^*$ which we also show in Fig. 2 (upper panels).

There are several notable features in Fig. 2 data. Focusing on Figs. 2(b) and (e), or (c) and (f), we note that $E_2^*m^*$ and $g^*m^*$ are increasingly enhanced over their band values as $r_s$ is increased, as expected in an interacting electron picture. The numerical values of $E_2^*m^*$ and $g^*m^*$ at different spin and valley are close despite the fact that they represent the system's response to very different external stimuli: $E_2^*m^*$ measures the rate of valley polarization with strain while $g^*m^*$ is the rate of spin polarization as a function of applied magnetic field. This observation suggests the similarity between spin and valley as two discrete degrees of freedom. However, when we combine the measurements of $E_2^*m^*$ and $g^*m^*$ with the corresponding $m^*$ data and deduce the values of $E_2^*$ and $g^*$ for different spin and valley occupations (Figs. 2(a) and (d)), the contrast between spin and valley becomes apparent. For $v_2$ (Fig. 2(a)), $E_2^*$ is enhanced over the band value and shows a slight increase with $r_s$ but does not show much dependence on spin
subband occupation. In contrast, for $s_2$, Fig. 2(d) reveals that $g^*$ has a strong
dependence on the valley occupation: although $g^*$ is enhanced over the band
value and increases with $r_s$, $g^*$ for $v_1$ is much larger than
it is for $v_2$. We highlight a noteworthy feature of Fig. 2 data: $m^*$, $g^*$, and $E_2^*$ reveal a clear contrast between spin and valley degrees of freedom while $g^*m^*$ and $E_2^*m^*$ do not. It appears as if $m^*$, $g^*$ and $E_2^*$ conspire to make the susceptibilities $g^*m^*$ and $E_2^*m^*$ behave similarly!

Theoretically, if spin and valley are considered only as discrete degrees
of freedom, then they are indistinguishable. Why is this not so in our 2DES?
In AlAs 2DES the spin and the valley indices are not
identical. The two valleys in wide AlAs quantum wells have $anisotropic$
Fermi contours whose major axes are rotated by 90 degrees with respect to each other. Therefore, the interaction
between electrons that have the same valley but different spin index
might be different from that between electrons that have the
same spin but different valley index. To examine this possibility, we describe here experimental data in two other 2DESs, namely those confined to either a narrow AlAs quantum well (well width $<$ 5 nm) or to a Si-MOSFET. In a \textit{narrow} AlAs quantum well, the electrons occupy a single valley with its major axis pointing out of plane and an in-plane \textit{isotropic} Fermi contour \cite{ShayeganPSS06}, while in a Si-MOSFET they occupy two such valleys \cite{ShashkinPRL03}.

The data for these two systems, taken from Refs. \cite{ShashkinPRB02,VakiliPRL04,GokmenPRB09} are summarized in the right panels of Fig. 2. Note that these data correspond to $s_2$ and should be compared to the data shown in the central panel of Fig. 2. Such comparison reveals that overall trends are qualitatively similar. In particular, in the Si-MOSFET case (Fig. 2(i)) where we have $v_2$, enhancements of $g^*m^*$ and $m^*$ track each other so that the deduced $g^*$ appears only slightly enhanced and its enhancement has a very weak dependence on $r_s$ (Fig. 2(g)). This is very similar to what is seen in Fig. 2(d) for the wide AlAs $v_2$ case. In the narrow AlAs quantum well (Fig. 2(h)) where we have $v_1$, on the other hand, the $g^*m^*$ enhancement is much larger than the $m^*$ enhancement and it grows faster with $r_s$. The deduced $g^*$ therefore exhibits a significant and $r_s$ dependent enhancement (Fig. 2(g)), similar to the $v_1$ case for the wide AlAs quantum well (Fig. 2(d)). We conclude that the contrast between the valley and spin degrees of freedom is not because of the Fermi contour anisotropy and might have a more intrinsic origin.

There are other non-ideal factors such as finite layer thickness and disorder which can give non-universal
corrections to the renormalization of $m^*$ (and susceptibilities) \cite{DePaloPRL2005, AsgariSSC2004, ZhangPRB05}. Finite layer thickness softens the Coulomb interaction but cannot cause a difference
between spin and valley degrees of freedom. In our measurements, we apply parallel magnetic field ($B_{\parallel}$) to fully spin polarize the 2DES or to tune the (Landau) energy levels. In a 2DES with finite electron layer thickness, $B_{\parallel}$
couples to the orbital motion of the electrons and leads to an
increase of $m^*$ \cite{GokmenPRB08}. However, because of the very small
electron layer thickness in our AlAs samples ($\leq$ 15 nm), we expect that this increase is less than 5\% even at $B_{\parallel} = 15$ T.

As for disorder, its effect of on $m^*$ has been studied theoretically \cite{AsgariSSC2004}, and it has been concluded that $m^*$ is larger when impurity scattering is taken into account compared to a clean system. We speculate that the difference between the valley and spin we observe might come from the differences in scattering mechanisms between states with opposite
spins or valleys. A scattering event requires the conservation of total spin and momentum. An electron
scattering from one valley to another requires a large momentum
transfer because the valleys are located near the edges of the Brillouin
zone. However, an electron scattering from one spin to another
within the same valley requires a small momentum transfer (on the
order of the Fermi wave-vector) and some magnetic impurity to conserve
the total spin. It is possible that these scattering
mechanisms are different in the presence of interaction and
disorder. An understanding of the contrast between spin and valley degrees of freedom in $m^*$, $g^*$ and $E_2^*$ renormalization awaits future theoretical developments.

We thank the NSF and DOE for support.

\break

\end{document}